\documentclass{article}
 \usepackage{amssymb,latexsym,amsfonts}
 \usepackage[dvipdf]{epsfig}
 \usepackage{color}

 \def\be{\begin{equation}}
 \def\ee#1{\label{#1}\end{equation}}

\begin{document}

 \title{\bf Cosmological model with non-minimally coupled fermionic field}

 \author{M. O. Ribas$^{1,2}$,  F. P. Devecchi$^1$ and G. M. Kremer$^1$ \\
              $^1$Departamento de F\'\i sica,
Universidade Federal do Paran\'a, Curitiba, Brazil\\
                $^2$Faculdades Integradas Esp\'\i rita, Curitiba, Brazil}
 \date{}

 \maketitle


 \begin{abstract}

 A model for the Universe is proposed whose constituents are: (a) a
dark energy field modeled by a fermionic field non-minimally coupled
with the gravitational field, (b) a matter field which consists of
pressureless baryonic and dark matter fields and (c) a field  which
represents the radiation and the neutrinos. The coupled system of
Dirac's equations and Einstein field equations is solved numerically
by considering a  spatially flat homogeneous and isotropic Universe.
It is shown that the proposed model can reproduce the expected
red-shift behaviors of the deceleration parameter, of the density
parameters of each constituent and of the luminosity distance.
Furthermore, for small values of the red-shift the constant which
couples the fermionic and gravitational fields has a remarkable
influence on the density and deceleration parameters.
 \end{abstract}

\section{Introduction}

In the last decade several works appeared in the literature with the
aim the explanation of the astronomical observations \cite{1} that
the Universe is presently expanding with a positive acceleration.
The great number of the works admits the existence of a component of
the Universe -- the so-called dark energy -- which is the
responsible for the present acceleration of the Universe, but whose
nature still remains unknown. The most natural candidate for dark
energy was initially a cosmological constant (see e.g. \cite{2}),
but other models for dark energy were also proposed. The most common models, also
called quintessence models,  make use of scalar fields and
barotropic equations of state where the ratio between the pressure
and energy density assumes negative values (e.g. ~\cite{3} and the
references therein). Other models consider that the dark energy
field is described by exotic equations of state like the Chaplygin
fluid~\cite{4} or the van der Waals fluid~\cite{5}.

Some cosmological models were also investigated in the literature by
considering fermions as sources of the gravitational field \cite{6}.
Recently, the authors~\cite{7} proposed a cosmological model in a
dissipative Universe and showed that the fermions could be the
responsible for accelerated regimes in the early Universe by
simulating an inflaton field and in the present epoch by simulating
 a dark energy field. Another paper on this subject was published
 very recently \cite{8}.

In most of the works which investigate the role of the scalar field
as source of the gravitational field it is common to consider it to
be minimally coupled to the gravitational field. However, there
exist some works in the literature \cite{9} where it was
investigated the effects of non-minimally coupled scalar fields
during the evolution of the Universe.

The aim of the present work is to describe a spatially flat,
homogeneous and isotropic Universe characterized by the
Robertson-Walker metric and whose constituents are a dark energy
field, a matter field and a field which represents the radiation and
the neutrinos -- from now on called radiation field. The dark energy
field is modeled by a fermionic field which is non-minimally coupled
with the gravitational field whereas the  matter field represents
the pressureless baryonic and dark matter fields. The radiation  and
matter fields are supposed to behave as non-interacting fields. The
basic field equations for the evolution of the Universe follow from
Dirac's equations and Einstein's field equations which are solved
numerically for given initial conditions. The red-shift behaviors of
the deceleration parameter, of the density parameters of each
constituent and of the luminosity distance, which result from the
numerical solutions, are compared with the available data set in
order to verify the viability of the proposed model.

The manuscript is structured as follows: in Sec. II the Dirac
equations and Einstein field equations are derived from the action
of a non-minimally coupled fermionic field. The field equations for
a flat, homogeneous and isotropic Universe are given in Sec. III,
which are rewritten in terms of the red-shift in Sec. IV. In Sec. V
the cosmological solutions are obtained by solving numerically the
system of field equations for given initial conditions. We close the
work with Sec. VI where we address to some remarks and sum up the
results obtained. The metric signature used is  $(+,-,-,-)$ and
units have been chosen so that $8\pi G=c=\hbar=1$.

\section{Non-minimally coupled fermionic field}

We are interested in investigate a  Universe modeled by a mixture
whose constituents are the fields of dark energy, matter and
radiation. As was pointed out in the previous section, the dark
energy is described by a fermionic field which is non-minimally
coupled with the gravitational field, whereas the fields of matter
and radiation are considered as non-interacting. The action for this
model reads
 \be
 S=\int \!\sqrt{-g}\,d^4 x
 \left[{1\over2}(1-\xi\overline\psi\psi)R+{\cal L}_f+{\cal L}_{m}+{\cal L}_r\right],
 \ee{1}
 where ${\cal L}_{m}$ and ${\cal L}_{r}$ denote the
 Lagrangian densities of  matter and radiation, respectively.
 Furthermore, $\psi$ and $\overline{\psi}$ are
 the spinor field and its adjoint, respectively, and $\xi$ is the
 coupling constant between the spinor field and the curvature
 scalar $R$.  The spinor fields are treated here as classically commuting
 fields~\cite{7,10}. Finally, ${\cal L}_{f}$ is the Lagrangian
 density of the fermionic field which for  massless fermions reads
 \be
 {\cal L}_f=\frac{\imath}{2}\left[ \overline\psi\,\Gamma^\mu
D_\mu\psi-(D_\mu\overline\psi)\Gamma^\mu\psi\right] -V.
 \ee{2}
In eq. (\ref{2}) $V$ -- which is only a function of $\psi$ and
$\overline\psi$ --  describes the potential density of
self-interaction between fermions. Moreover, $\Gamma
^{\mu}=e^\mu_a\gamma^a$ are the generalized Dirac-Pauli matrices,
$e^a_\mu$ denotes the tetrad or ``vierbein" while the covariant
derivatives are given by
 \be
 D_\mu\psi=\partial_\mu\psi-\Omega_\mu\psi,\qquad
D_\mu\overline\psi=\partial_\mu\overline\psi+\overline\psi\Omega_\mu.
 \ee{3}
Above, the spin connection $\Omega_\mu$ is defined by
 \be
 \Omega_\mu=-\frac{1}{4}g_{\rho\sigma}[\Gamma_{\mu\delta}^\rho-e_b^\rho
 \partial_\mu e_\delta^b]\Gamma^\sigma \Gamma^\delta,
 \ee{4}
 with $\Gamma^\nu_{\sigma\lambda}$ denoting the Christoffel symbols.

 From  Euler-Lagrange equations applied to the total Lagrangian density in eq. (\ref{1})
 it follows the Dirac equations for the spinor field  and its adjoint coupled
 to the gravitational field, namely
 \be
 \imath\Gamma^\mu D_\mu\psi-{dV\over d{\overline\psi}}-{\xi\over2} R\psi=0,
 \ee{5}
 \be
  \imath D_\mu\overline\psi\,\Gamma^\mu+{dV\over d\psi}+{\xi\over2}
 R\overline\psi= 0.
 \ee{6}

 The variation of the action (\ref{1}) with respect to the tetrad
 leads to Einstein's field equations
 \be
 R_{\mu\nu}-\frac{1}{2}g_{\mu\nu}R=-{T_{\mu\nu}\over 1-\xi\overline\psi\psi},
 \ee{7}
 where  $T_{\mu\nu} $ is the total energy-momentum tensor of the sources which
 is a sum of the energy-momentum tensor of the fields of fermions
 $T^{\mu\nu}_f$, matter $T^{\mu\nu}_{m}$
 and radiation $T^{\mu\nu}_{r}$, i.e.,
$T^{\mu\nu}=T^{\mu\nu}_f+T^{\mu\nu}_{m}+T^{\mu\nu}_{r}$.

The symmetric form of the energy-momentum tensor of the fermions
which follows from the variation of the action with respect to the
tetrad  reads
 $$
 T^{\mu\nu}_f=\frac{\imath}{2}\left[\overline\psi\Gamma^{(\mu}
 D^{\nu)}\psi -D^{(\mu}\overline\psi \Gamma^{\nu)}\psi\right] -g^{\mu\nu}{\cal L}_f
 $$
 \be
-\xi\left({\cal T}^{\mu\nu}
 -{{\cal T}^\sigma}_\sigma
 g^{\mu\nu}\right).
 \ee{8}
Above, the parentheses around the indices denote the symmetric part
of a tensor and it was introduced the tensor ${\cal T}_{\mu\nu}$
defined by
$$
 {\cal T}_{\mu\nu}=\left[D_{(\mu} D_{\nu)}\overline\psi \right]\psi
 +\overline\psi\left[D_{(\mu} D_{\nu)}\psi\right]
 $$
 \be
  +2D_{(\mu}\overline\psi D_{\nu)}\psi -\Gamma_{\mu\nu}^\sigma(D_\sigma\overline\psi)\psi
  -\Gamma_{\mu\nu}^\sigma\overline\psi(D_\sigma\psi).
 \ee{9}

\section{Field equations for Robertson-Walker metric}

From now on we shall investigate  cosmological solutions which can
be obtained  from the above proposed model when applied to a
homogeneous, isotropic and spatially flat Universe described by the
Robertson-Walker metric $ds^2=dt^2-a(t)^2\delta_{ij}dx^idx^j$, where
$a(t)$ denotes the cosmic scale factor.

For the Robertson-Walker metric the components of the tetrad, spin
connection and Dirac-Pauli matrices reduce to
 \be
 e_0^\mu=\delta_0^\mu,\quad
 e_i^\mu=\frac{1}{a(t)}\delta_i^\mu,\quad
 \Omega_0=0,\quad  \Omega_i=\frac{1}{2}\dot
a(t)\gamma^i\gamma^0,
 \ee{10}
 \be
\Gamma^0=\gamma^0,\quad  \Gamma^i=\frac{1}{a(t)}\gamma^i, \quad
\Gamma^5=-\imath\sqrt{-g}\,\Gamma^0\Gamma^1\Gamma^2\Gamma^3=\gamma^5.
 \ee{11}
 Moreover, in this case  the spinor field  is an exclusive function of time and
 the Dirac equations (\ref{5}) and (\ref{6}) become
 \be
 \dot\psi+{3\over2}H\psi+\imath\gamma^0{dV\over
 d{\overline\psi}}+\imath{\xi\over2}R\gamma^0\psi=0,
 \ee{12}
 \be
 \dot{\overline{\psi}}+{3\over2}H\overline\psi-\imath{dV\over
 d{\psi}}\gamma^0-\imath{\xi\over2}R\overline{\psi}\gamma^0=0,
 \ee{13}
 where $H=\dot a(t)/a(t)$ denotes the Hubble parameter and the
 dot represents a differentiation with respect to time.

 For a homogeneous and isotropic Universe the total energy-momentum tensor of the sources
 of the gravitational field  is written as
 $(T^\mu_\nu)=$ diag $(\rho,-p,-p,-p)$, where $\rho$ and $p$ denote the total
 energy density and pressure of the sources, respectively. In terms
 of the energy densities and pressures of the constituents, the
 total energy and pressure are written as
 \be
 \rho=\rho_f+\rho_{m}+\rho_r,\qquad
 p=p_f+p_{m}+p_r,
 \ee{14}
 i.e., as a sum of the energy densities and pressures of the
 fermionic, matter  and radiation fields, respectively.

 From Einstein's field equations (\ref{7}) it follows the modified forms of Friedmann and
 acceleration equations
 \be
 H^2={\rho\over3(1-\xi\overline\psi\psi)},\qquad {\ddot a \over a}=-{\rho+3p\over6(1-\xi\overline\psi\psi)},
 \ee{15}
 respectively.

 By differentiating the Friedmann equation  (\ref{15})$_1$ with respect to time  and taking into
 account: (i) the acceleration equation (\ref{15})$_2$, (ii) the Dirac equations
 (\ref{12}) and  (\ref{13}) and (iii)  that the self-interaction
 potential $V$ between the fermions does depend only on the bilinear $\Psi=\overline \psi\psi$,
 it follows the evolution equation for the total energy density of the sources
 of the gravitational field, namely
 \be
 \dot\rho+3H(\rho+p)=-{\rho\xi\dot\Psi\over(1-\xi\Psi)}.
 \ee{16}
 The above equation follows also from the Bianchi identities
 applied to the Einstein field equation (\ref{7}). Note that due to the coupling
 between the fermionic and the gravitational fields,
  the right-hand side of (\ref{16}) is not zero and represents the energy transfer between  them.
 As it is well known, only two among the three equations
 given in (\ref{15}) and (\ref{16}) are linearly independent.

 By assuming that the energy-momentum tensors of the constituents
 have the same form as the one for the total energy-momentum tensor,
 it follows from (\ref{8}) and (\ref{9}) that the energy density and pressure of the fermionic field
 read
 \be
 \rho_f=V+3\xi H\dot\Psi,
 \ee{17}
 \be
 p_f={dV\over d{\psi}}{{\psi}\over2}+{{\overline\psi}\over2}{dV\over
 d{\overline\psi}}-V+{\xi\over2}R\Psi
 -\xi(\ddot\Psi+2H\dot\Psi).
 \ee{18}
  Note that the existence of a coupling
 between the fermionic and the gravitational fields leads to a dependence of the
 energy density and pressure of the fermionic field with respect to  the energy density and pressure
 of all constituents through the Hubble parameter $H$ and the scalar curvature $R$.

 After the matter and radiation decoupling both fields behave as non-interacting
 fields so that the evolution equations for their  energy densities read
 \be
 \dot\rho_m+3H\rho_m=0,\qquad
 \dot\rho_r+4H\rho_r=0,
 \ee{19}
 by taking into account that matter is considered as a pressureless fluid,
 i.e., $p_m=0$ whereas the radiation obeys the barotropic equation
 of state $p_r=\rho_r/3$.

 The evolution equation for the energy density of the fermionic field
  follows from  (\ref{14}), (\ref{16}) and  (\ref{19}), yielding
 \be
 \dot\rho_f+3H(\rho_f+p_f)=-{\rho\xi\dot\Psi\over(1-\xi\Psi)},
 \ee{20}
 where the right-hand side of the above equation represents the energy
transfer of the fermionic field to the gravitational field due to
their coupling.

\section{Field Equations in terms of the red-shift}

In order to get   cosmological solutions for the proposed model, the
red-shift will be used as a variable instead of time
 thanks to the following relationships
 \be
  z={1\over a}-1,\qquad {d\over dt}=-H(1+z){d\over dz}.
  \ee{21}

 As a first step we note that that eqs. (\ref{19}) can be easily integrated leading to the
 well-known dependence of the energy densities of the matter and
 radiation fields with the red-shift
 \be
 \rho_m(z)=\rho_m(0)(1+z)^3,\qquad
 \rho_r(z)=\rho_r(0)(1+z)^4,
 \ee{22}
 where $\rho_m(0)$ and $\rho_r(0)$ are the values of the energy
 densities at present time $z=0$.

 Afterwards,  the energy density and the pressure of
 the fermionic field are expressed in terms of the red-shift,
 yielding
 \be
 \rho_f=V-3\xi H^2(1+z)\Psi^\prime,
 \ee{23}
$$
 p_f=\left[{2+\xi\Psi\over2(1-\xi\Psi)}\right]^{-1}\left\{{d V\over
 d\Psi}\Psi-V-\xi[(1+z)^2H^2\Psi^{\prime\prime}\right.
$$
 \be\left.
 -(1+z)H^2\Psi^\prime+(1+z)^2HH^\prime\Psi^\prime]+{\xi\Psi(\rho_m+\rho_f)\over2(1-\xi\Psi)}\right\},
 \ee{24}
 where the prime denotes the differentiation with respect to the
 red-shift $z$ and again by assuming that the self-interaction potential
 $V$ between the fermions does depend only on the bilinear $\Psi=\overline
 \psi\psi$. For a non-coupled fermionic field its energy density and
 pressure reduce to
 \be
 \rho_f=V,\qquad p_f={d V\over d\Psi}\Psi-V.
 \ee{25}

 Finally, a system of two coupled differential
 equations for the bilinear $\Psi$  and for the Hubble parameter $H$ is  obtained from the Dirac
 equations (\ref{12}), (\ref{13}) and from the Friedmann and
 acceleration equations (\ref{15}),  that reads
 \be
 (1+z)H\Psi^{\prime\prime}-[2H-(1+z)H^\prime]\Psi^\prime-3H^\prime\Psi=0,
 \ee{26}
 \be
  2H(1+z)(1-\xi\Psi)H^\prime=\rho+p.
 \ee{27}

 For the determination of the cosmological solutions from the above equations we have also to
 specify how the self-interaction potential
  between the fermions $V$ depend on the bilinear $\Psi$. Here we
  suppose that it is given by the power law $V=\kappa/[\Psi(z)]^\alpha$, where
  $\kappa$ and $\alpha$ are two free parameters.

\section{Cosmological solutions}

 For the  search of numerical solutions of the system of differential equations
 (\ref{26}) and (\ref{27}) together with (\ref{22}) through (\ref{24})
 we have to specify initial conditions for $\Psi(z)$,
 $\Psi^\prime(z)$ and $H(z)$. Here we shall suppose that in the
 radiation dominated era at $z_0=10^4$ (say) the value of the bilinear
  is very small $\Psi(z_0)=10^{-5}$ with a smaller value for its
 derivative with respect to the red-shift $\Psi^\prime(z_0)=10^{-13}$
 so that the  fermionic field rolls very smoothly
 during the evolution of the Universe. Moreover, we shall adopt
 the following  values for  the density parameters $\Omega_i(z)=\rho_i(z)/\rho(z)$ at present
 time: $ \Omega_{f}(0)=0.72,$ $\Omega_{m}(0)=0.279916$ and $\Omega_{r}(0)=8.4\times10^{-5}$
 (see  e.g.~\cite{11}). Hence the initial value for  the
 dimensionless Hubble parameter can be obtained from Friedmann
 equation (\ref{15})$_1$ together with (\ref{22}) and (\ref{23}),
 yielding
 \be
 H(z_0)\!=\!\sqrt{{\Omega_m(0)(1+z)^3+\Omega_r(0)(1+z)^4+
 \overline\kappa/[\Psi(z_0)]^\alpha\over3(1-\xi\Psi(z_0))+3\xi(1+z_0)\Psi^\prime(z_0)}},
 \ee{28}
 where $\overline\kappa$ is also a dimensionless quantity.

 To find numerical solutions of the system of differential equations (\ref{26}) and
 (\ref{27}), still remains to fix values for the free parameters: (a) $\overline\kappa$ and
 $\alpha$ of the self-interaction potential between the fermions and
 (b)  the coupling constant between the fermionic field and the
 gravitational field $\xi$.  The following values  $\overline\kappa=0.6$ and
 $\alpha=1/10$ were found by imposing that the present value of the density
 parameter of the fermionic field matches the expected value $
 \Omega_{f}(0)=0.72$ when there is no coupling, i.e., when $\xi=0$.
 The search for values of these parameters by changing the initial conditions is
 exhaustible, specially  due to the fact
 that the system of differential equations (\ref{26}) and (\ref{27})
 is very unstable. Here we shall adopt the above quoted  values.

 \begin{figure}
\begin{center}
\includegraphics[width=6.5cm]{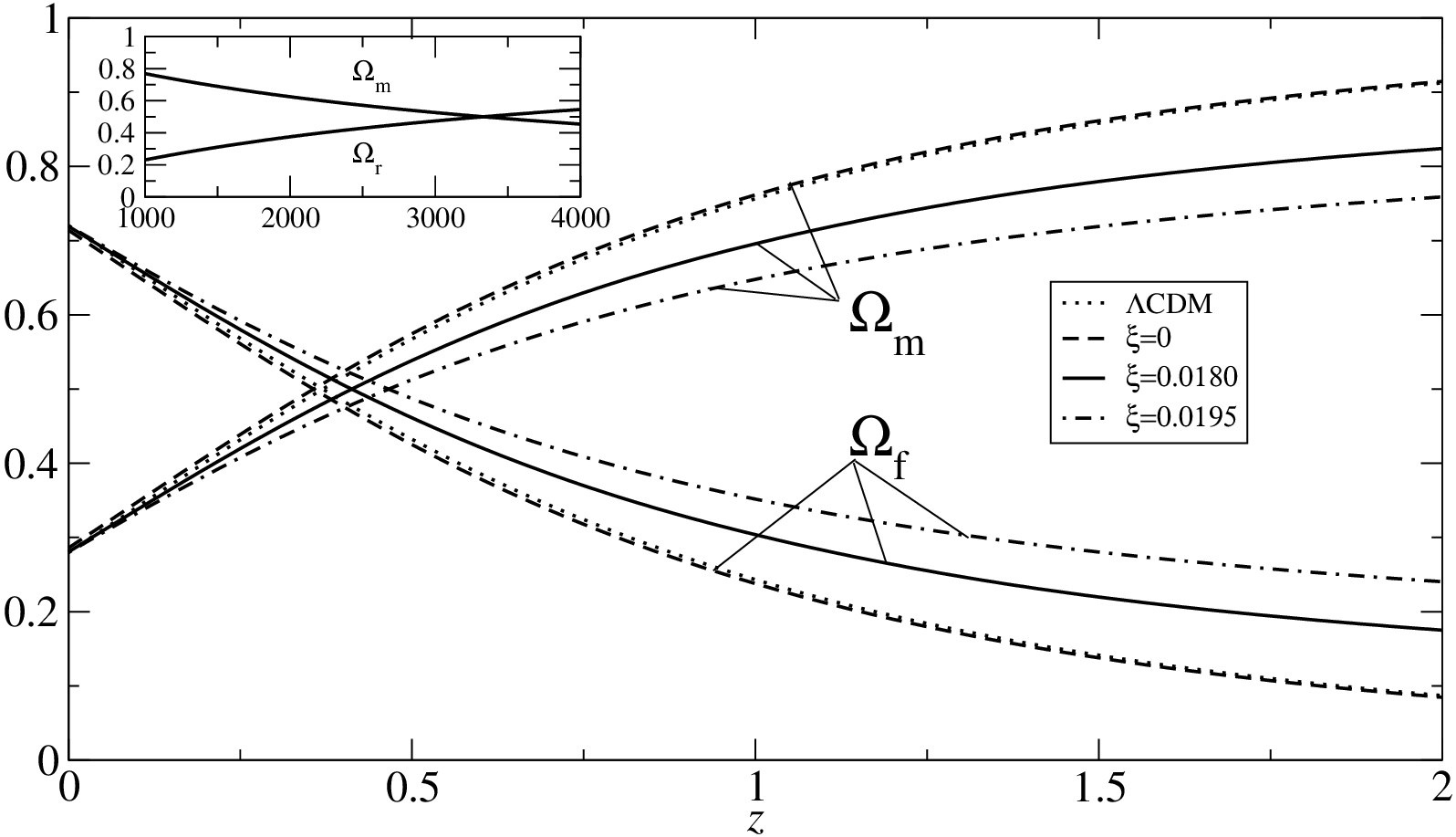}
\caption{Density parameters as functions of the red-shift for the
$\Lambda$CDM model and for the fermionic model with different values
of $\xi$. Large frame $0\leq z\leq2$, small frame $1000\leq
z\leq4000$.}
\end{center}
\end{figure}

 In  figure 1 the density parameters are plotted as functions of the
 red-shift in the range between $0\leq z\leq2$ (large frame) and in the
 range between $1000\leq z\leq4000$ (small frame).
 We infer from the large frame of this figure that the coupling parameter $\xi$
 has influence on the
 decay of the dark energy field and the corresponding increase of
 the matter field. Moreover, the density parameter of the field of
 radiation in this range is very small.
 When $\xi=0$ there is no sensible difference
 between the $\Lambda$CDM model and the model where the fermions
 represent the dark energy field and there is no coupling. We
 note that by increasing the coupling parameter $\xi$ the
 decrease of the fermionic field and the corresponding increase of the matter
 field are less accentuated. This fact can be understood by noting that
 according to eq. (\ref{20}) a part of the energy density of the fermionic field is transferred
 to the gravitational field due to its coupling. Furthermore, one can observe that the system of differential
 equations is very sensible to small changes of the coupling
 parameter. The density parameter of the fermionic field becomes
 negligible when  $z\approx5$ for the case when there is no coupling ($\xi=0$) and
 when $z\approx200$ for  the other two cases $\xi=0.0180$ and $\xi= 0.0195$, with
 a more accentuated decay of the former  with respect to the latter.
 The density parameters of the  radiation and matter fields are
 represented in the small frame in figure 1, which shows that
 the equality between the matter and radiation occur at
 $z\approx3300$. It is noteworthy that here we have
 taken into account the degrees of freedom of the relativistic
 neutrinos.

\begin{figure}
\begin{center}
\includegraphics[width=6.5cm]{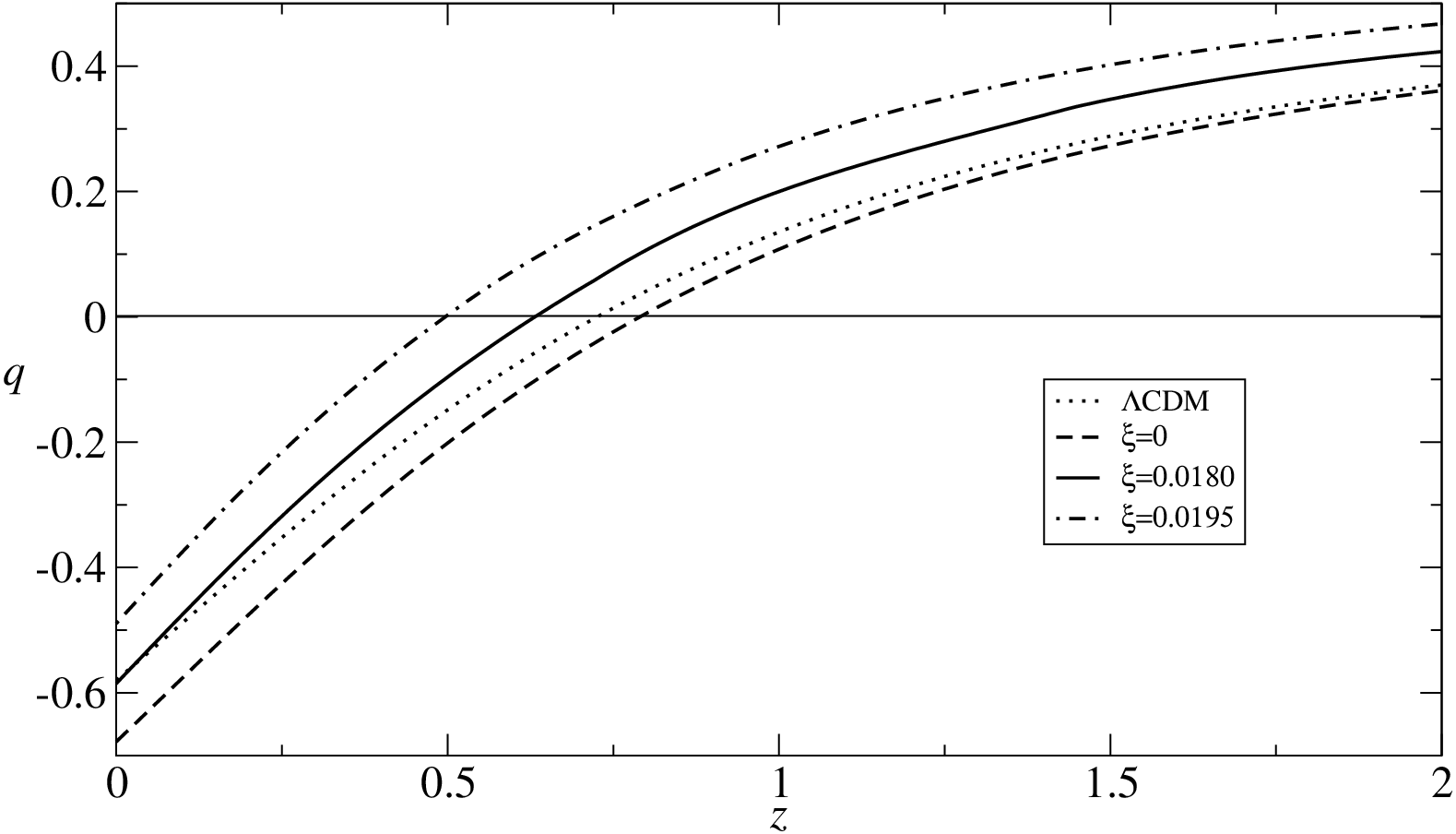}
\caption{Deceleration parameters as functions of the red-shift for
the $\Lambda$CDM model and for the fermionic model with different
values of $\xi$.}
\end{center}
\end{figure}

 The deceleration parameter $q=1/2+3p/2\rho$ is plotted in figure 2
 as function of the red-shift. One can infer  that the coupling
 parameter $\xi$ has also a very remarkable  influence on the deceleration
 parameter $q$. It is important to note that the $\Lambda$CDM model
 furnishes a different solution with respect to the uncoupled fermionic model
 ($\xi=0$). Indeed, although the values of $q$ coincide at large values of the red-shift
 ($z>1.5$)  their present values at $z=0$ and  red-shifts $z_t$ that correspond to
 the transition from a decelerated  to an accelerated regime read:
 (a) $\Lambda$CDM model $q(0)=-0.58$ and $z_t=0.73$ and (b)
 uncoupled model $q(0)=-0.68$ and $z_t=0.80$. The corresponding
 values for $\xi\neq 0$ are: (c) $q(0)=-0.58$ and $z_t=0.64$ for
 $\xi=0.0180$ and (d) $q(0)=-0.49$ and $z_t=0.50$ for
 $\xi=0.0195$. Apart the last case all others fit more or less the experimental
 values for the deceleration parameter, namely, $q(0)=-0.74\pm0.18$ (see \cite{12}) and $z_t=0.46\pm0.13$
 (see \cite{13}).

 In figure 3 it is plotted the difference between the apparent $m$
 and the absolute magnitude $M$ of a source, denoted by $\mu_0$ and defined by
 \be
 \mu_0=m-M=5\log \left\{(1+z)\int_0^z{dz'\over
 H(z')}\right\}+25,
 \ee{29}
 where the quantity between braces represents the luminosity distance in Mpc. The
 circles in this figure correspond to the experimental data for super-novae of type Ia taken from
 the recent work by Riess et al. \cite{14}.
 It is important to call attention to the fact
 that the values of the coupling constant $\xi$ chosen above do not lead
 to sensible differences in the values of $\mu_0$ so that only one curve for the fermionic model is represented
 in the figure. One can infer from the small frame of figure 3 that there is a very good agreement of the curves
 of the $\Lambda$CDM model and of the fermionic model
 with the experimental data for red-shifts in the range  $0\leq
 z\leq0.5$. From the large frame of this figure one observes
 that there exist some differences between the curves of the two models at higher
 red-shifts $0.5\leq z\leq1$.

\begin{figure}
\begin{center}
\includegraphics[width=6.5cm]{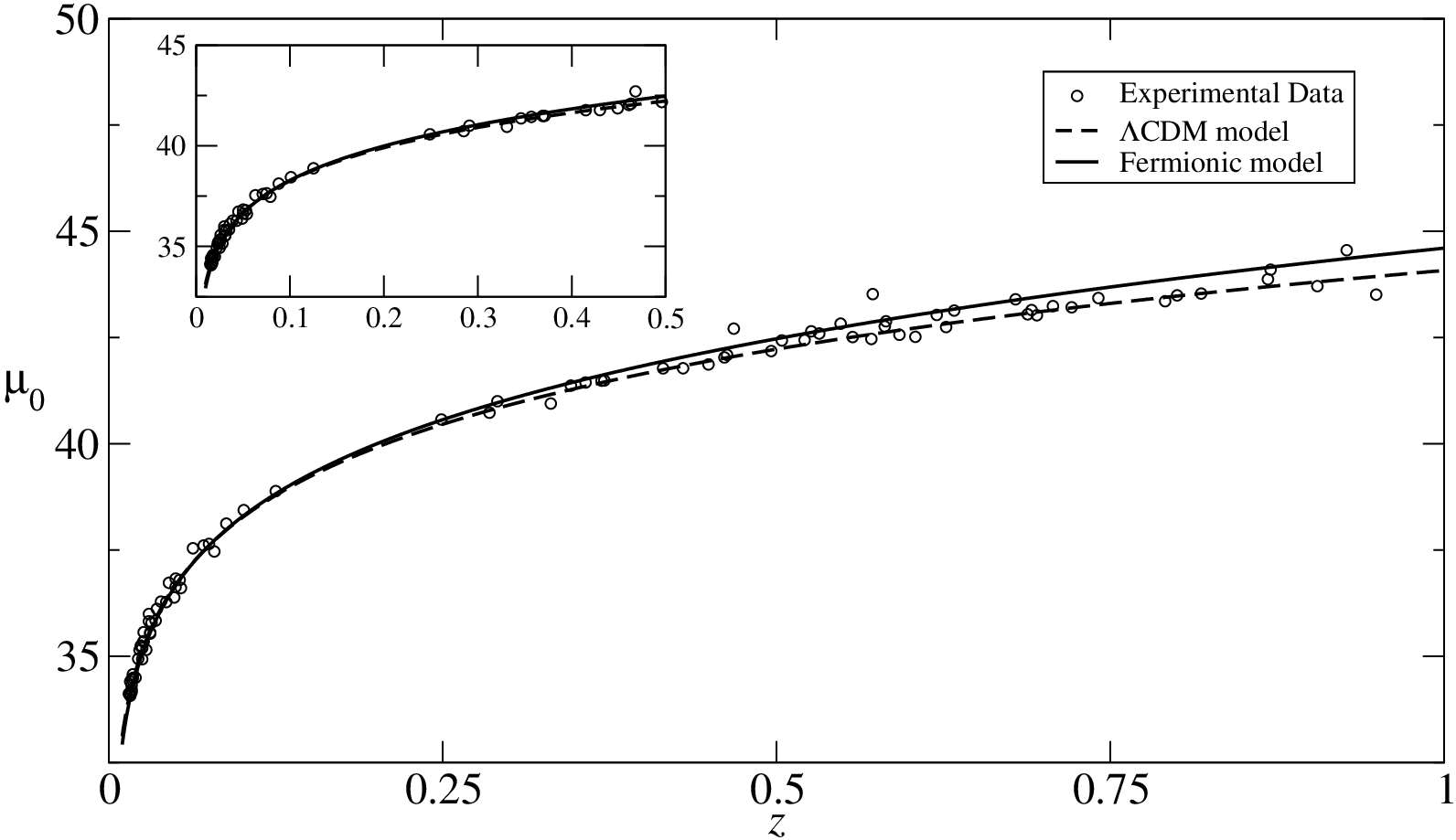}
\caption{$\mu_0$ as function of the red-shift $z$ for the
$\Lambda$CDM model and for the fermionic model. Large frame $0\leq
z\leq1$, small frame $0\leq z\leq0.5$.}
\end{center}
\end{figure}

\section{Final remarks and conclusions}

In this work we have considered a classical fermionic field non-minimally coupled with the gravitational field.
Classical spinors were discussed by Armend\'ariz-Pic\'on and Greene in Ref. [6] and for more details one is referred to this work. Here we point out that: (i)  the expectation value of a spinorial field in a physical state is a complex number and not a Grassmannian number and (ii) the spinorial field can be treated classically if its state is
close to the vacuum.

Recently, exact cosmological solutions for a Universe described by a Robertson-Walker metric
and modeled by a neutral scalar field and neutrinos were obtained in the work ~\cite{15}. The model of this work
is different from the model presented here since in the former  the neutrinos are considered
as a Majorana fermion field while the neutral scalar field represents the acceleron of the mass-varying neutrinos.

One important point which deserves attention is the one concerning
the equivalence of the above description with the one in which the
Einstein's field equations is written as
 \be
 R_{\mu\nu}-\frac{1}{2}g_{\mu\nu}R=-{\tilde T_{\mu\nu}},
 \ee{30}
 where  $\tilde T_{\mu\nu}=\tilde
 T^{\mu\nu}_f+T^{\mu\nu}_{m}+T^{\mu\nu}_{r}$, with
 $$
 \tilde T^{\mu\nu}_f=\frac{\imath}{2}\left[\overline\psi\Gamma^{(\mu}
 D^{\nu)}\psi -D^{(\mu}\overline\psi \Gamma^{\nu)}\psi\right] -g^{\mu\nu}{\cal L}_f
 $$
 \be
-\xi\left({\cal T}^{\mu\nu}
 -{{\cal T}^\sigma}_\sigma
 g^{\mu\nu}\right)-\xi\overline\psi \psi\left(R_{\mu\nu}-\frac{1}{2}g_{\mu\nu}R\right).
 \ee{31}

In this  case we have $\tilde {T^{\mu\nu}}_{;\nu}=0$  thanks to the
Bianchi identities. If we  write the total energy-momentum tensor as
$(\tilde T^\mu_\nu)=$ diag $(\tilde\rho,-\tilde p,-\tilde p,-\tilde
p)$ -- so that $\tilde \rho$ and $\tilde p$ represent the total
energy density and the total pressure, respectively  -- it follows
 \be
 H^2={\tilde\rho\over3},\qquad {\ddot a\over a}=-{\tilde\rho+3\tilde p\over6},
 \qquad\dot{\tilde\rho}+3H(\tilde\rho+\tilde p)=0,
 \ee{32}
 which are the usual form of the Friedmann, acceleration and energy
 conservation equations, respectively.

 Equations (\ref{32}) are not in contradiction with the corresponding eqs. (\ref{15}) and
 (\ref{16}), since one can prove that the energy density and pressure of the latter case
 are related with those of former case by
 \be
 \tilde\rho={\rho\over(1-\xi\Psi)},\qquad
 \tilde p={p\over(1-\xi\Psi)}.
 \ee{33}
 Hence, it is straightforward to obtain eqs. (\ref{15}) and (\ref{16}) from eqs.
 (\ref{32}) and (\ref{33}) by the use of Dirac's equations (\ref{12}) and
 (\ref{13}).

 Another remark is that the non-minimally coupling
 is important only at small  red-shifts and its effect
 becomes negligible at high red-shifts where the matter-radiation
 decouples and the nucleosynthesis begins. Moreover, the method
 adopted here was to solve the system of differential equations
 (\ref{26}) and (\ref{27}) by specifying small values for
 $\Psi(z_0)$ and $\Psi'(z_0)$ at a high red-shift $z_0$ so that the fermionic field
 rolls very smoothly during the evolution of the Universe. Another
 method follows by combining equations (\ref{12}) and (\ref{13}) so
 that the resulting equation leads to
 \be
 {\Psi(z)\over\Psi(z_0)}={(1+z)^3\over(1+z_0)^3}.
 \ee{34}
From the above equation one can obtain that the energy density of
 the fermionic field reads
 \be
 \rho_f=\left[{K\over[\Psi(z)]^n}-{3\xi[\rho_m(z)+\rho_r(z)]\Psi(z)\over[1-\xi\Psi(z)]}\right]
 \left[{1-\xi\Psi(z)\over1+2\xi\Psi(z)}\right].
 \ee{35}
 Now by specifying an initial condition to the bilinear $\Psi(z)$ one
 can made a similar analysis on the behavior of the density parameters, of the
 deceleration parameter and of the luminosity distance as was done
 in Sec. V.

 {As a last remark we call attention to the fact that we have considered a very simple type self-interaction
 potential which depends only on the bilinear $\overline\psi\psi$.
 With this choice, the fermionic field seems
 to behave as a bosonic field, but a comparison of the field equations and
 solutions of the present work with those obtained by Binder and Kremer in Ref. ~\cite{9} for a bosonic field non-minimally coupled with the gravitational field shows that the solutions are quite different.  A more general
 self-interaction potential should also depend on the  pseudo-scalar invariant
 $ (i\overline {\psi }\gamma ^5 \psi)^2$ as the one investigated by the authors in Ref. \cite{7}.
 This work is in progress and shall appear in a forthcoming paper.}

 To sum up: (a) the proposed model is very sensible
 concerning the choice of the initial conditions and the choice
 of the free parameters $\xi$, $\overline \kappa$ and $\alpha$;
 (b) the density parameter of the fermionic and matter fields decrease more
 slowly with respect to the red-shift  when the fermions are coupled  ($\xi\neq0$)
 due to  a more efficient energy transfer between  the fermionic and the
 gravitational fields;  (c) by increasing the coupling constant $\xi$
 the transition from a decelerated to an accelerated regime
 occurs at earlier red-shifts and (d) there is no significant
 difference in parameter $\mu_0$ -- which is related to the
  luminosity distance -- for coupled and uncoupled fermionic
 fields.

\section*{Acknowledgments}

The authors FPD and GMK acknowledge the support by
Conselho Nacional de Desenvolvimento Cient\'\i fico e Tecnol\'ogico (CNPq).


\end{document}